\documentclass[journal]{vgtc}                     %

\onlineid{0}

\vgtccategory{Theoretical \& Empirical}

\vgtcpapertype{evaluation}

\title{The Arrangement of Marks Impacts Afforded Messages:\\ Ordering, Partitioning, Spacing, and Coloring in Bar Charts}

\author{
  \authororcid{Racquel Fygenson}{0000-0002-0705-9000},
  \authororcid{Steven Franconeri}{0000-0001-5244-9764}, and 
  \authororcid{Enrico Bertini}{0000-0002-9932-0551}
}

\authorfooter{
  \item
  	Racquel Fygenson and Enrico Bertini are with Northeastern University.
  	E-mail: fygenson.r | e.bertini @northeastern.edu
  \item
  	Steven Franconeri is with Northwestern University.
  	E-mail: franconeri@northwestern.edu.
}

\abstract{%
Data visualizations present a massive number of potential messages to an observer. One might notice that one group's average is larger than another's, or that a difference in values is smaller than a difference between two others, or any of a combinatorial explosion of other possibilities. The message that a viewer tends to notice – the message that a visualization ‘affords’ – is strongly affected by how values are arranged in a chart, e.g., how the values are colored or positioned. Although understanding the mapping between a chart’s arrangement and what viewers tend to notice is critical for creating guidelines and recommendation systems, current empirical work is insufficient to lay out clear rules. We present a set of empirical evaluations of how different messages--including ranking, grouping, and part-to-whole relationships--are afforded by variations in ordering, partitioning, spacing, and coloring of values, within the ubiquitous case study of bar graphs. In doing so, we introduce a quantitative method that is easily scalable, reviewable, and replicable, laying groundwork for further investigation of the effects of arrangement on message affordances across other visualizations and tasks. Pre-registration and all supplemental materials are available at \url{https://osf.io/np3q7} and \url{https://osf.io/bvy95}, respectively.
}

\keywords{Perception \& cognition, Methodologies, Human-subjects qualitative studies, Human-subjects quantitative studies, Charts, diagrams and plots, General public}

\teaser{
  \centering
  \includegraphics[width=\linewidth]{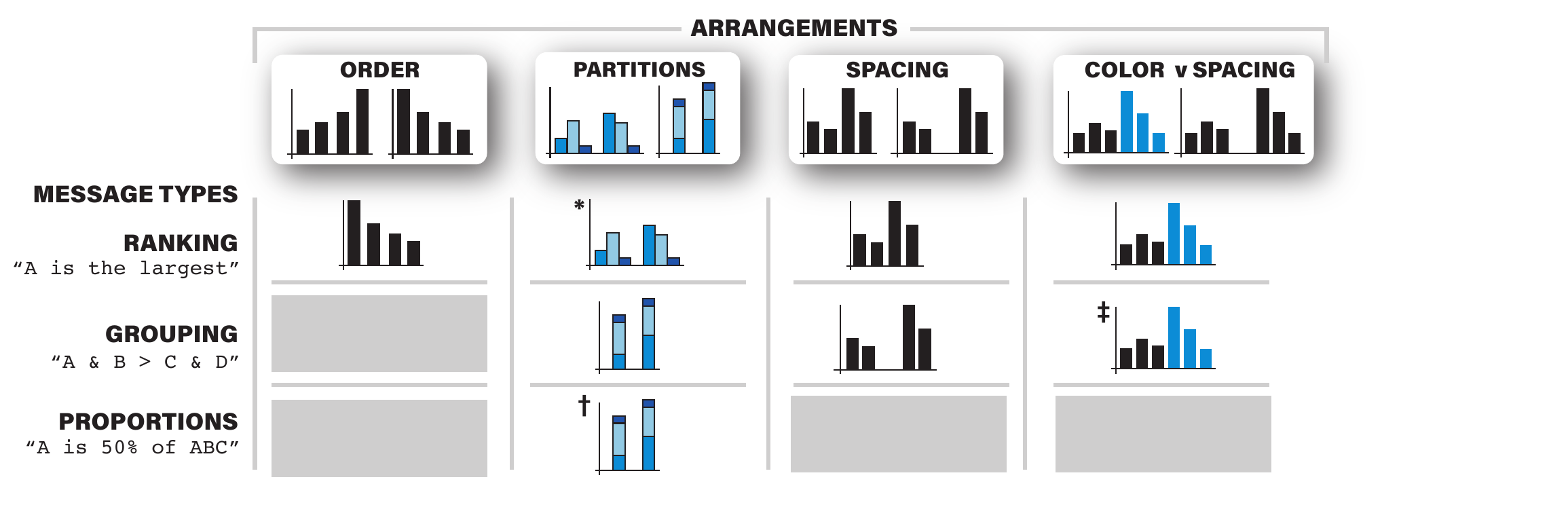}
  \caption{A sample of presented findings: each cell shows the best option we found. Column headers show the options we compared, and  row headers show tested message types. Gray rectangles indicate that the corresponding message-arrangement was not studied. Of tested arrangements, $\ast$ is best when comparing individual parts and $\dag$  is best when 3 colors are present -- see \cref{fig:exp2results}. $\ddag$ is best when 2 colors are present -- see 
  \cref{fig:exp4results}.
  }
  \label{fig:teaser}
}

\graphicspath{{figs/}{figures/}{pictures/}{images/}{./}} %

\usepackage{tabu}                      %
\usepackage{booktabs}                  %
\usepackage{lipsum}                    %
\usepackage{mwe}                       %
\usepackage{makecell}
\usepackage{wrapfig}

\usepackage{mathptmx}                  %

\begin{document}

\firstsection{Introduction}

\maketitle

Visualization evaluation and design are often guided by a  ranking of visual variables developed on precision-based criteria (e.g., response time, exactness of read values) \cite{munzner2015visualization, WareInfoVis, north-measuring-insight, isenberg-sys-review-vis-techniques, lam-sys-review-vis-techniques}. Other visualization guidance is based off intuition \cite{tufte-vis-display-quant}, or extrapolated from cognitive psychology experiments that use far simpler stimuli (e.g., sets of shapes) and different participant tasks \cite{Wertheimer-1923-gestalt, oyama-1961-proximity, oyama-1999-similarity-motion, benav-similarity-proximity, hochberg-silverstein-brightness-proximity, quinlan-wilton-prox-vs-similarity}.
 
\begin{figure}[t]
    \includegraphics[width=0.49\textwidth]{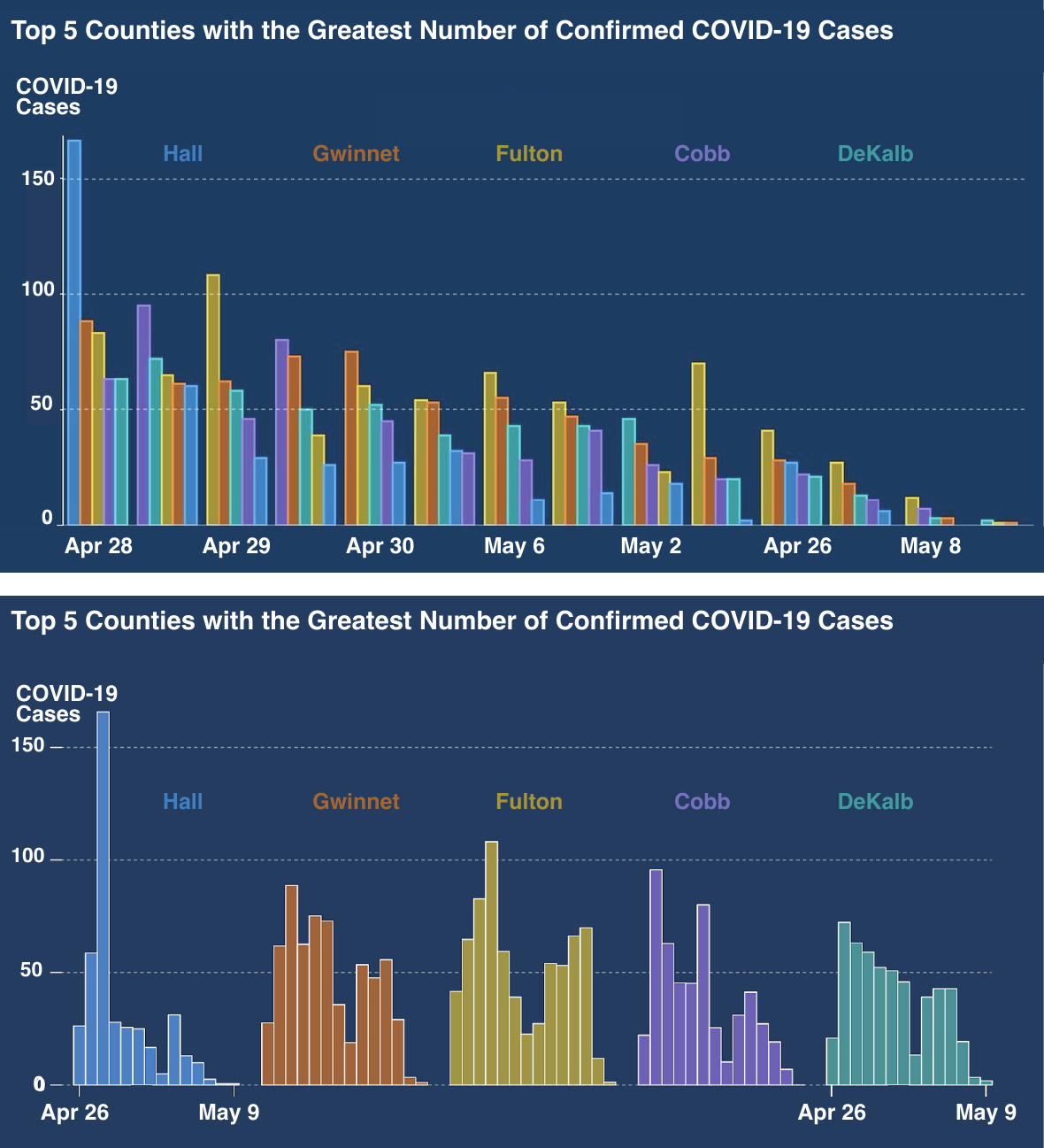}
    \caption{These charts show the same information with differently ordered and grouped bars. What patterns are most obviously communicated by the top chart? Are these same patterns obvious in the bottom? Recreated from \cite{cairobars}.}
    \label{fig:notallscatter}
    \vspace{-5mm}
\end{figure}

Precision-based evaluation provides limited guidance in designing an effective visualization. While precision can ensure quantitative data is estimated accurately from graphical depictions, it is not sufficient to guarantee efficacy. 
Visualization designs can convey data in precise ways, yet not make an intended message obvious, and imprecise designs can still make intended messages obvious and intuitive \cite{bertini-all-chart-not-scatterplot}. 
Similarly, two visualizations can show the same data with equal precision, but communicate significantly different messages.
Consider the pair of graphs in \Cref{fig:notallscatter},  taken from Alberto Cairo's popular blog ``The Functional Art.'' Both graphs encode number of COVID-19 cases using the height of aligned bars, but differences in ordering and spatial proximity of their bars convey markedly different trends in case numbers. The top graph sorts bars in descending order regardless of time, implying a consistently decreasing trend, while the bottom communicates that the numbers of cases by county increase before they decline. Thus, it is possible for simple changes in the arrangement of parts of a chart to impact the message that a viewer is likely to grasp. More generally, as existing research posits, data visualizations' design can afford potential takeaways \cite{xiong-reasoning-affordances, xiong-visual-arrangement-of-bars, zacks-tversky-bars-lines, shah-graph-comprehension}. In practice, past research has investigated afforded messages by examining how differences in visualizations can compel viewers to reason differently \cite{xiong-reasoning-affordances}, alter the type of comparisons they make \cite{xiong-visual-arrangement-of-bars}, and most commonly vary their description of underlying information \cite{zacks-tversky-bars-lines, shah-graph-comprehension, xiong-visual-arrangement-of-bars}.

In this paper, we explore a novel metric for evaluating afforded takeaways: 
Do some arrangements of marks (i.e., visual objects in a graph) make messages more obvious than others? To answer this question, we need to 1) enumerate a possible set of mark arrangements and a possible set of subsequently afforded messages and 2) investigate if these arrangements impact the obviousness of the identified messages. 

We present four experiments that investigate how four arrangements of data marks (ordering, partitioning, spacing, coloring) affect the subjective match of visualization designs to a set of messages (ranking judgments, group comparisons, and part-to-whole relationship judgments).

\section{Related Work}
\label{sec:relwork}

 While precision-centric evaluation methods remain extremely popular\cite{isenberg-sys-review-vis-techniques}, alternate methodologies 
 have been advocated in a collection of evaluation-focused papers \cite{wall-vis-evaluation, plaisant-vis-evaluation, north-measuring-insight, bertini-beyond-time-error-beliv, bertini-all-chart-not-scatterplot}, and motivated the long-running IEEE VIS workshop \textit{BEyond time and errors: novel evaLuation methods for Information Visualization} (BELIV). In alignment with growing consideration for new metrics of evaluation, studies have explored visual metaphors \cite{ziemkiewicz-visual-metaphor}, memorability \cite{bateman-chartjunk, borkin-memorability, borgo-memorability}, deeper insights \cite{north-measuring-insight}, implicit takeaways\cite{xiong-visual-arrangement-of-bars}, and afforded reasoning\cite{xiong-reasoning-affordances}.

\subsection{Current Methods for Exploring Visualization\\ Affordances}
\label{sec:2.1}
In the late 1990s, Zacks and Tversky, and Shah et al. explored differences in reader takeaways incurred by bar and line charts, finding that bars imbue a sense of discreteness, while line charts imply continuous relationships\cite{zacks-tversky-bars-lines, shah-graph-comprehension}. In both of these seminal works, researchers employed qualitative methodologies, showing graphs to participants and then hand-coding their open-ended descriptions. This method provides strong benefits by allowing findings to arise organically, without the need for pre-declared hypotheses. The inverse of this method--in which researchers describe a relationship between data points and ask participants to draw a corresponding graph--offers similarly beneficial evidence \cite{zacks-tversky-bars-lines, walny-data-sketches, gaba-comparison-design}.

Hand-coding qualitative survey responses continues to be used to study visualization affordances, by asking participants to type out or voice their takeaways.  \cite{xiong-reasoning-affordances, xiong-visual-arrangement-of-bars, xiong-metadata, lee-sensemaking}. But this methodology is time-consuming and labor-intensive. Even more problematic, as diligently reported by Xiong et.al, this hand-coding is often unable to resolve the natural syntactic and semantic ambiguities in sentences that people type \cite{xiong-visual-arrangement-of-bars}. As a simple example, imagine a bar graph showing the sizes of two birds and two squirrels. If a viewer says, "the birds are bigger than the squirrels", they could mean that any bird is bigger than any squirrel, or that the birds are bigger than squirrels on average. Open-ended methods are powerful exploratory tools, but require, the sometimes impossible, resolving of ambiguities to effectively study visualization affordances. 
Thus, we present a complementary methodology to such approaches. We employ a confirmatory design that restricts the space of tested stimuli and messages, but provides efficient, replicable data to verify the impact of visual arrangements on afforded messages. 

Similar approaches include asking participants to report their opinion, often using Likert scales\cite{south-likert-vis-eval}, on how much different visualizations support semantic variables (e.g., "stable", "rigid", "complete")\cite{ziemkiewicz-implied-dynamics}, on the trustworthiness or bias of visualizations \cite{padilla-mfv-trust, kong-2019-trustrecall}, on their agreement with provided statements\cite{kong-2018-slantstitles, kong-2019-trustrecall}, or on amount of risk to themselves or others before and after seeing visualizations showing pandemic information\cite{padilla-covid-riskperception}.

We present a comparably empirical approach, but focus on general reader takeaways, a subject matter that--to the best of our knowledge--has only been quantitatively evaluated once before, in Xiong et al.'s Experiment 2, and with a much smaller (n=45 experts vs our n $\geq$ 130 general public) sample\cite{xiong-visual-arrangement-of-bars}. For a more detailed comparison of our work to Xiong et. al, see SM7 in Supplemental Materials.

\subsection{Bar Chart Research}\label{sec:relwork-barchart}
Bar charts, one of the most prevalent types of visualization\cite{lee-VLAT}, are a common subject of visualization evaluations, and produce study results that have been generalised to other graph types\cite{munzner2015visualization}. Foundational research in visualization, including the widely cited Cleveland \& McGill, Zacks \& Tversky, Shah et al, Bateman et al, and Heer \& Bostock papers seek to evaluate fundamental paradigms of visualizations and their communication by studying bar charts \cite{cleveland-mcgill, zacks-tversky-bars-lines, bateman-chartjunk, heer-bostock-repr, talbot-bars, zacks-bar-precision-depth-cues}. Takeaways from these papers establish core tenets of bar chart interpretation, including how accurately one can discern bar chart lengths given different placement and heights of the bars \cite{cleveland-mcgill, heer-bostock-repr, talbot-bars, zacks-bar-precision-depth-cues}, and that arranging bars in groups using irregular spacing leads to readers "visual[ly] chunking" their takeaways accordingly\cite{shah-graph-comprehension}. For the most part, best practices using bar charts can be informed via the effectiveness ranking of channels\cite{munzner2015visualization}.

In this work, we explore the effect of bar chart arrangements beyond classic manipulated variables (e.g., x-axis alignment, height differences) and the classic dependent variable of precision (e.g., response time, read accuracy). To further the understanding of visualization design, we include conditions that have been tested (e.g., bar alignments vs misalignment), and those that have yet to be investigated (e.g., spacing vs coloring to convey grouping of bars), and focus on the impact of these conditions on message obviousness. Thus we present novel results on how bar chart arrangements' obviousness of messages can align, or fail to align, with precision-based design decisions. See Figures \hyperref[fig:exp1results]{4}, \hyperref[fig:exp2results]{5}, \hyperref[fig:exp3results]{6}, and
\hyperref[fig:exp4results]{7} for tested messages, bar charts, and results.

\section{Study Rationale}
\label{rationale}
We start by identifying arrangements of marks that may have an influence on afforded messages (e.g., formats that compel viewers to compare data, focus on trends, or guide a specific reading sequence). We prioritize arrangements that can generalize to multiple graphical representations and therefore have an impact beyond bar charts. We avoid studying arrangements inherent to a specific visual mark (e.g., density/texture in choropleths which does not have an equivalent in line charts). We also aim to limit use of arrangements that require multiple visual channels to encode information. In this study, we test
\textbf{Ordering}, \textbf{Partitioning}, \textbf{Spacing}, and \textbf{Coloring}.

\textbf{Ordering} describes the sequence in which objects can be arranged in a visualization. Examples of ordering in different visualizations include the sequence of bars in a bar chart, segments in a pie chart, columns and rows in a matrix visualization, and plots in a small multiple visualization.

\textbf{Partitioning} pertains to the division of a graphical mark into sub-parts, and those sub-parts' corresponding placement in a visualization. Partitioning can be found in pie charts, treemaps, and stacked bar charts.

\textbf{Spacing}, in this context, addresses the use of spatial proximity to organize visual objects into groups. Spacing is used in grouped bar charts, exploded pie charts, and grouping within Sankey diagrams or alluvial plots.

\textbf{Coloring}, in this context, describes the use of different hues or levels of saturation to group elements and/or distinguish between elements of different types. Examples of coloring used in this way include colored dots in scatter plots, colored bars in bars charts, colored rectangles in treemaps, and colored lines in multi-line charts.

We hypothesize that these four variations of arranging marks influence the strength of different messages extracted from visualizations. Specifically, we identify the following types of afforded messages and their influencing arrangements:
\textbf{Ranking} (Ordering),
\textbf{Grouping} (Spacing, Partitioning, Coloring), and 
\textbf{Part-to-whole relationships} (Partitioning).
We provide details of how these versions are compared and measured experimentally in the \hyperref[sec:m&m]{Materials \& Methods} section.

\subsection{Experimental Approach}
\label{sec:expapp}
The following is a high-level overview of the motivation behind our experimental approach. For experimental design details, see the \hyperref[sec:m&m]{ Materials \& Methods} section. 

Our main experimental goal is to explore how the arrangements outlined in the previous subsection influence readers’ interpretation of visualizations. 

As detailed in \hyperref[sec:2.1]{Section 2.1}, we seek to complement popular open-ended methods of studying visualization affordances with the following, more structured approach:
\begin{enumerate}
    \item Identify message types that may be afforded by visualizations.
    \item For each message type (e.g., ranking, grouping), develop example messages to test
    \item Using previous research, knowledge, or open-ended experimental methods, hypothesize which variations in mark arrangements may strengthen or weaken the affordance of test messages.
    \item Design visualizations that contain the arrangements hypothesized to be a good match for each message type, along with visualizations that are hypothesized to not match well.
    \item Showing one message at a time, ask participants to select which of the visualization designs best matches the message. We do so in practice by asking participants which visualization makes the message “the most obvious” to them.
    \item Quantify the strength of the arrangement-to-message fit according to the frequency with which participants select tested visualizations.
    \item Determine support, or lack thereof, for hypotheses by evaluating the proportion of participants that report each visualization as making the tested message the most obvious.
\end{enumerate}
This approach has complementary advantages and disadvantages to the qualitative approach of collecting readers’ interpretations through open-ended questions. Our approach makes message matching more quantifiable, less noisy and more accessible to reviewing and replication. On the other hand, it is solely confirmatory, hinging upon researchers’ choice of tested messages and graphic variations, and is potentially influenced by nuances in message wording. 

Fortunately, partnering our approach with an open-ended exploratory method (see \cref{sec:2.1}) can mitigate the first limitation. The second limitation can be addressed, as in \hyperref[fig:exp2results]{Experiment 2}, by testing messages with the same meaning but slight re-wordings to investigate if nuanced wording is confounding results. In this study, we have found rewording to have no effect.

\section{Materials \& Methods}
\label{sec:m&m}
In this work, we examine the arrangements of marks in bar charts, and how they impact the obviousness of afforded messages. We conduct four separate, confirmatory, within-subjects studies in which we study the effects of ordering, partitioning, coloring, and irregularly spacing bars (see top row of \cref{fig:exp1results,fig:exp2results,fig:exp3results,fig:exp4results}). In contrast with the majority of current research on visualization affordances \cite{shah-graph-comprehension, yang-explaining-w-examples, zacks-tversky-bars-lines, gaba-comparison-design, Tversky_Ordering}, we employ a quantitative methodology, in which research participants select one of four shown graphs that makes a given message the most obvious to them. In doing so, we reduce the uncertainty around confirming and replicating qualitative experiments, but also reduce the investigative scope of our experiment; our experimental conclusions, and thus our hypotheses, are context-specific. Accordingly, any hypothesis that message M will be made the most obvious by graph G1, must be qualified with the context that M is only made more obvious by G1 than G2, G3, or G4. Our pre-registered hypotheses, experimental design, and analysis plan are available at \url{https://osf.io/np3q7}.

\subsection{Investigative Questions \& Hypotheses}
\label{sec:hyps}
\subsubsection{Experiment 1 - Order}
In Experiment 1, we investigate the effect of sorting bars on messages concerning rank (see \Cref{fig:exp1results}). While some work has established written language influences mental ordering schema~\cite{Tversky_Ordering, dickinson-ordering-bias}, and other research has hypothesized about the cognitive effort required to identify extrema given variously ordered bar charts\cite{schwartz-perceptual-effort-sorted-bars}, we are unaware of any studies that explore differences in ascending and descending graphs' interpretation.

We hypothesize that \textbf{(H1A)} bar charts arranged in descending order from left to right (\cref{fig:exp1results}, C) will make messages about first-, second-, and third-largest bars (\cref{fig:exp1results}, Ordering.1, .2, .3)  the most obvious. Conversely, bar charts arranged in ascending order from left to right (\cref{fig:exp1results}, D) will make messages about first-, second- and third- smallest bars (\cref{fig:exp1results}, Ordering.4, .5, .6) the most obvious. 

This hypothesis stems from previous cognitive psychological research that shows left-to-right visual scanning associations stemming from left-to-right languages influence mental ordering schema~\cite{Tversky_Ordering, dickinson-ordering-bias}. Because all of our participants speak English fluently and currently reside in the United States, we hypothesize that they are pre-disposed to reading bar charts from left-to-right, and thus any ordering-specific messages would be made most obvious by chart arrangements in which the extreme associated with the ordering in question is further towards the left.

\subsubsection{Experiment 2 - Partitions}
Experiment 2 investigates the arrangement of bars that encode part-to-whole data. Specifically, this experiment studies how bars that are placed side-by-side afford proportion-specific and comparison messages more or less strongly than those that are stacked vertically (see \Cref{fig:exp2results}). Previous studies establish that axis-aligned bars will be more precisely interpreted than those that are not\cite{cleveland-mcgill, talbot-bars, heer-bostock-repr}. Talbot et al. further establish that vertically aligned, adjacent bars are more likely to be mis-estimated to add up to 100\%--and thus considered a whole--than vertically aligned, spatially separated bars \cite{talbot-bars}. Building on Talbot et al.'s discovery, we investigate stacked and un-stacked bars' effect on the obviousness of messages concerning comparison between parts, comparison between wholes, and the existence of proportional relationships, in an effort to see if affordances of these messages align with previously accepted paradigms of stacked bar charts. 

We hypothesize \textbf{(H2A)} stacked bar charts (\cref{fig:exp2results}, C \& D) will make messages about the the whole of parts (e.g., comparison among summed parts, \cref{fig:exp2results}, Partitions.1, .2) more obvious than side-by-side bar charts (\cref{fig:exp2results},  A \& B). We also hypothesize \textbf{(H2B, H2D)} stacked bar charts (\cref{fig:exp2results},  C \& D) will make messages about individual parts as a proportion of a whole (e.g., clip sales in the West make up 50\% of all clip sales, \cref{fig:exp2results}, Partitions.7, .8, .9, .10) more obvious than side-by-side bar charts (\cref{fig:exp2results},  A \& B). These hypotheses are informed by theory on physical-visual metaphors \cite{ware-visual-thinking-for-design}, and past research that indicates that stacked bars imply wholeness and a part-to-whole relationship more strongly than side-by-side layouts\cite{talbot-bars}. We also hypothesize \textbf{(H2C)} side-by-side bar chart arrangements (\cref{fig:exp2results},  A \& B) will make messages specific to the identification and comparison of individual parts in a part-to-whole visualization (e.g., clip sales in the West vs clips sales in the East, (\cref{fig:exp2results},  Partitions.3, .4, .5, .6) more obvious, via the same, albeit inverse, reasoning as our previous two hypotheses.

\subsubsection{Experiment 3 - Spacing}
Experiments 3 and 4 investigate the affordance of grouping messages, given bars with varied ordering, spatial proximity, and coloring. 
Experiment 3 tests uniform vs irregular spacing to determine if, as is currently maintained in visualization\cite{shah-graph-comprehension, munzner2015visualization}, visual perception\cite{century-gestalt-vis-perception}, and psychology\cite{brooks-chapter-perceptual-grouping, Wertheimer-1923-gestalt} literature, increasing the space between bars--and therefore the proximity of some bars to others--affords grouping.

We hypothesize that \textbf{(H3A)} bar charts with uniform spacing (\cref{fig:exp3results}, A \& B) will make messages about overall extrema more obvious than bar charts with grouping implied via irregular spacing (\cref{fig:exp3results}, C \& D). This hypothesis stems from pilot study results and, while logical given gestalt principles\cite{century-gestalt-vis-perception}, was not immediately obvious to us before collecting pilot data. We also hypothesize that \textbf{(H3B)} bar charts with irregular spacing defining elements in groups (\cref{fig:exp3results}, C \& D) will make messages that discuss those groups more obvious than bar charts with uniform spacing (\cref{fig:exp3results}, A \& B).

\subsubsection{Experiment 4 - Color vs Spacing}
Experiment 4 compares the strength of spatial grouping to color grouping in bar charts (see \Cref{fig:exp4results}). Research on the hierarchy of visual grouping mechanisms has found that proximity conveys grouping more strongly that similar coloring\cite{quinlan-wilton-prox-vs-similarity, benav-similarity-proximity, munzner2015visualization, brooks-chapter-perceptual-grouping, franconeri-what-works-vis-data-comm}. Interestingly, the majority of studies that investigate visual grouping do not examine bar charts, instead, focusing on dot lattices\cite{quinlan-wilton-prox-vs-similarity, benav-similarity-proximity, hochberg-silverstein-brightness-proximity, oyama-1961-proximity, oyama-1999-similarity-motion, brooks-chapter-perceptual-grouping}. 

We hypothesize that \textbf{(H4A)} bar charts with color groups and regular spacing (\cref{fig:exp4results}, A \& B) will make messages about overall extrema more obvious than those with groups defined by spacing (\cref{fig:exp4results}, C \& D), because the communication of grouping will be less strong in the color-grouped charts and thus easier to ignore when evaluating extrema over multiple groups. Informed by the same known hierarchy, we also hypothesize that \textbf{(H4B)} bar charts with groups defined by spacing (\cref{fig:exp4results}, C \& D) will make messages that discuss those groups more obvious than charts with groups defined by color (\cref{fig:exp4results}, A \& B). 

\subsubsection{Methodological Checks}
Lastly, to check the methodological rigor of our survey, we include the following message-graph questions and their pre-registered hypotheses. To confirm that participants are not swayed by familiarity bias, thus always reporting that height-ordered bar charts (\cref{fig:exp1results}, C \& D) make all messages most obvious, we hypothesize that \textbf{(H1C)} charts with bars sorted in a specific order (\cref{fig:exp1results}, A) will make messages comparing bars grouped in that order (\cref{fig:exp1results}, Order.7) most obvious due to the proximity of the bars in questions. 

In Experiment 2, we also test some pairs of messages with alternate wordings and sentence structures to examine the effect of the style of messages on our results. We hypothesize that \textbf{(H2E)} messages communicating the same concept, despite rewording, (\cref{fig:exp2results}, Partitions.3 and .4, .7 and .8, .9 and .10) will not produce different results.

\subsection{Stimuli Design}
Experiments 1-4 investigate variations in bar chart arrangements (see \cref{fig:exp1results,fig:exp2results,fig:exp3results,fig:exp4results}). Unlike much prior affordance work in visualization, we focus on varying arrangements within bar charts as opposed to visualization encoding types (e.g. bar charts, line charts, pie charts) in the interest of evaluating design decisions that are less commonly investigated in visualization literature, education, and recommendation systems. This decision also reinforces the validity of our survey design; asking participants to select between multiple types of visualizations increases the risk of familiarity bias (i.e., participants always selecting visualizations that they have seen more often) clouding participant judgement. Due to the lack of visual variance, we believe the familiarity differential of bar charts with varying arrangements (e.g., ordered ascending vs descending) is much smaller than that plausible of different visualization types (e.g., pie chart vs tree-map). 

\subsubsection{Experiment 1 - Order}
Experiment 1 investigates ordering of bars and is motivated by a lack of research into the effects of such design decisions. The majority of research on ordering of bars generally fails to investigate higher-level, participant-reported takeaways in favor of response time, precision, eye-tracking, and cognitive effort models\cite{schwartz-perceptual-effort-sorted-bars, feeney-ordering-task-speed, dickinson-ordering-bias, michal-vis-routines}.

In all four experiments we test four different visualization conditions for the sake of methodological consistency. Experiment 1 consists of the same bar chart in ascending, descending, and alphabetical order, as well as a fourth, "wildcard" ordering in which tallest bars are centered forming a $\wedge$ shape (see \cref{fig:exp1results}). This last arrangement was motivated by the desire to test four conditions, and by an interest in how the Gestalt Law of Symmetry, which states that people tend to perceive symmetrical shapes and prefer visual symmetry\cite{Wertheimer-1923-gestalt, century-gestalt-vis-perception}, might have an unexpected effect on message obviousness (spoiler alert: it didn't).

\subsubsection{Experiment 2 - Partitions}
Experiment 2 investigates the representation of part-to-whole bar charts. The primary motivation behind the development of its visualization conditions was to explore the obviousness of part-to-whole relationships given different partitioning. The hierarchy of visual encoding channels (as discussed in the \hyperref[sec:relwork]{Related Work}) is universal in informing effective visualization design \cite{munzner2015visualization, WareInfoVis}, and can be used to justify the replacement of all part-to-whole visualizations (i.e., pie charts, stacked bars) with side-by-side bar charts. This replacement prioritizes precision but has not been shown to better facilitate the communication of relationships or other non-precision messages. In fact, previous work investigating the efficacy of pie and bar charts challenges the effectiveness hierarchy when completing certain tasks \cite{simkin-bar-v-pie}. Experiment 2 seeks to investigate how the hierarchy of effectiveness compares to the affordance of part-to-whole messages in side-by-side (aligned) and stacked (unaligned) bar charts. Thus, Experiment 2's visualization space consists of one dataset split into two groups of three bars, both side-by-side and stacked (\cref{fig:exp2results}, B \& D), and the same data split into three groups of two bars both side-by-side and stacked ( \cref{fig:exp2results}, A \& C).

To determine color scheme, we selected three colors from a widely used categorical color palette from Tableau\footnote{\url{https://help.tableau.com/current/pro/desktop/en-us/viewparts_marks_markproperties_color.htm}}, a popular software for making visualizations. We selected these colors by avoiding hues that are strongly associated with warning (i.e., red, orange, yellow). Next we used Color Oracle\footnote{\url{https://colororacle.org/index.html}}, free software that simulates common forms of Color Vision Deficiency (CVD)\cite{color-oracle-paper}, to evaluate and slightly alter the luminance of our chosen colors so as to increase their distinction for viewers with CVD.  

\subsubsection{Experiment 3 - Spacing}
Experiments 3 and 4 seek to investigate how color and spatial arrangements of marks afford grouping. Experiment 3 is designed to replicate previous findings that irregular spacing strongly implies groups among bar charts\cite{burns-autoapproach-for-grouped-bars}. Thus Experiment 3's stimuli design consists of two different orderings of bars, each regularly spaced (\cref{fig:exp3results}, A \& B), and then irregularly spaced into groups (a condition with two groups of three bars, and a condition with three groups of two bars (\cref{fig:exp3results}, C \& D).

\subsubsection{Experiment 4 - Color vs Spacing}
Experiment 4 shares much of the same motivation as Experiment 3, but investigates a less strongly supported theory on grouping in bar charts. While generally proximity is agreed to imply grouping more strongly than similar coloring\cite{quinlan-wilton-prox-vs-similarity, benav-similarity-proximity, munzner2015visualization, brooks-chapter-perceptual-grouping, franconeri-what-works-vis-data-comm}, this has not been directly measured in bar charts. Thus, Experiment 4 presents a novel investigation into the hierarchy of afforded grouping in bar charts. To do so, Experiment 4 replicates proximity grouped conditions from Experiment 3 (\cref{fig:exp4results}, C \& D), and compares them to equivalent bar charts that use color grouping instead (\cref{fig:exp4results}, A \& B). We reuse the CVD-friendly color scheme from Experiment 2 in this experiment as well.

\subsection{Procedure}
All four of our within-subjects experiments were implemented through a Qualtrics\footnote{\url{https://www.qualtrics.com/}} survey. After reading and approving a consent form, participants were given the option to self-report their education level and if they had CVD (mentioned by name and colloquialized as ``colorblindness" in our survey). Participants were then instructed to make their browser window as large as possible and primed on the types of graphs they would see (see SM1 in Supplementary Materials for language used). They were then shown a page comprised of four charts with varied arrangements, a short sentence describing the content of the charts, and, as an attention check, a message with a fill-in-the-blank drop-down consisting of two possible answers, one of which correctly described the data depicted in all four chart conditions. Participants were instructed to \textit{1. Use the charts below to fill in the blank.} and \textit{2. Then select the chart that makes the statement below most obvious to you.} See SM2 in Supplemental Materials for an example survey question. Participants were shown between 6 and 10 of these questions, depending on the number of tested messages in the experiment.

This methodology, which presents quantitative and easily replicable evidence of subjective takes, stands in contrast to many similarly motivated investigations, which show participants visualization stimuli and ask them to describe it \cite{zacks-tversky-bars-lines, shah-graph-comprehension, xiong-visual-arrangement-of-bars, yang-explaining-w-examples}, or show participants a description and ask them to represent the information with a visual creation\cite{zacks-tversky-bars-lines, gaba-comparison-design, Tversky_Ordering}. As mentioned in the \hyperref[sec:2.1]{Section 2.1}, to the best of our knowledge, the only experiment with similar methodology to ours is Xiong et al.'s Experiment 2 (see SM7 in Supplemental Materials for a more detailed comparison)\cite{xiong-visual-arrangement-of-bars}.

Both the order in which participants viewed questions, and the order charts were presented in the quadrant of every question were randomized using the Qualtrics "randomization" functionality. The order of the drop-down answers for the fill-in-the-blank was not randomized, but held consistent with terms like "smaller," "less," and "least" appearing above terms like "larger," "more," and "most," so as not to confuse participants or lead to incorrect selection despite correct comprehension.

We added the fill-in-the-blank  question as both an attention check, and to compel participants to actually read and consider the message when reporting the graph that made it the most obvious. Without this experimental design detail, we would have little way of knowing if participants actually read and reported their opinions on the message, because all four conditions show the same data and are therefore technically ``correct" answers. %
While our pre-registered analysis plan\footnote{\url{https://osf.io/np3q7}} dictates excluding a participant's chart selection if they incorrectly answer the corresponding drop-down question, we find very little inconsistency between reported obviousness of charts from participants who correctly and incorrectly answer the drop-down. See SM4 in Supplemental Materials for a comparison of results with and without this exclusion criteria. 

\subsection{Participants}
Participants were recruited via the online platform Prolific\footnote{\url{prolific.co}}. Prolific connects scientific researchers with eligible human studies participants, and offers a number of services to facilitate high-quality, ethical human-subjects research, including enacting specified inclusion and exclusion criteria, encouraging fair pay rates for participants, and facilitating compensation directly. Using Prolific, we recruited participants who were over the age of 18, fluent in English, current residents of the United States, and had high ($\geq 98\%$) approval rates on the platform, and constructed a study population that was roughly balanced on reported sex, as stated in our pre-registered study plan\footnote{\url{https://osf.io/np3q7}}. Also via Prolific, we compensated all participants 1.60USD for their participation, given an anticipated participation time of 8 minute, for an estimated rate of 12.00USD/hour.

\section{Results}
\label{sec:results}
\subsection{Participants}
A total of 610 participants were recruited via Prolific. Of these, 591 (Exp. 1 n = 147, Exp. 2 n = 166, Exp. 3 n = 140, Exp. 4 n = 138) completed the full survey with no higher than a 30\%  error rate, passing the universal exclusion criteria, and were included in our final data analysis. 
For a breakdown of participants' reported sex, education and color vision deficiency for each experiment, see \hyperref[tab:demo-participants]{Table 1}. Exact sample size per message varies based on number of participants who selected the corresponding drop-down correctly, although all sample sizes are equal to or more than our minimum pre-registered sample size of 128. For a breakdown of sample size per tested message see SM3 in Supplementary Materials.

 Initially, Experiment 1 included multiple un-piloted messages that resulted in very high ($>25\%$) error rates. We hypothesized that these errors were most likely due to ambiguous or overly convoluted messages. We re-wrote these messages to be more straightforward \footnote{for differences in the preliminary and final run of Experiment 1, compare the pre-registered design (\url{https://osf.io/np3q7}) with the design reported in this paper}, and re-ran the entire experiment, drastically decreasing error rates to $\leq 12\%$. We report the results from the final Experiment 1 in this paper.

\begin{table}[tb]
  \caption{Demographics of Participants per Experiment}
  \vspace{-3mm}
  \label{tab:demo-participants}
  \scriptsize%
	\centering%
  \begin{tabu}{%
	r%
	*{7}{c}%
	*{2}{r}%
	}
  \toprule
     & \rotatebox{0}{Exp. 1} &   \rotatebox{0}{Exp. 2} &   \rotatebox{0}{Exp. 3} &   \rotatebox{0}{Exp. 4} \\
  \midrule
  Female & 75 & 84 & 72 & 70 \\
  Male & 72 & 82 & 68 & 68 \\
  \midrule
  Some high school & 2 & 0 & 0 & 2 \\
  High school/GED & 32 & 48 & 35 & 27 \\
  \makecell{Tech/community college,\\associates degree } & 26 & 30 & 26 & 29 \\
  \makecell{Undergraduate degree} & 63 & 55 & 51 & 61 \\
  \makecell{Graduate degree} & 21 & 24 & 24 & 16 \\
  \makecell{Doctoral degree}& 3 & 8 & 4 & 3 \\
  \midrule
  Does not have CVD & 141 & 163 & 139 & 135\\
  Has CVD & 5 & 1 & 140 & 0 \\
  Did not answer & 1 & 2 & 0 & 3\\
  \textbf{n} & \textbf{144} & \textbf{166} & \textbf{140} & \textbf{138} \\
  \bottomrule
  \end{tabu}%
  \vspace{-6mm}
\end{table}

\subsection{Analysis}
 We exclude all participants who answered $> 30\%$ of all drop-down answers incorrectly. %

Participant responses are analyzed using the Sison-Glaz procedure for estimating multinomial proportion confidence intervals\cite{sison-glaz}, as implemented by the Python library statsmodels.stats \cite{statsmodels}. Due to the multinomial nature of this procedure, no correction for family-wise error rate is necessary. Using the worst-case multinomial proportion table (Table 1) from Steven K. Thompson's  ``Sample Size for Estimating Multinomial Proportions"\cite{thompson-sample-size}, we determine minimum sample size for a 95\% confidence interval within a maximum specified distance from the true proportion, \textit{d}, of 0.1 to be 128 participants per experiment. We elect to only conduct a visual analysis of the confidence intervals, avoiding null hypothesis significance testing and its common pitfalls (e.g. type II statistical errors)\cite{Dragicevic2016}. We present and discuss results of all four experiments using language and best practices of statistical analysis for Human Computer Interaction \cite{Dragicevic2016}.

\begin{figure}[t!]
\includegraphics[width=0.49\textwidth]{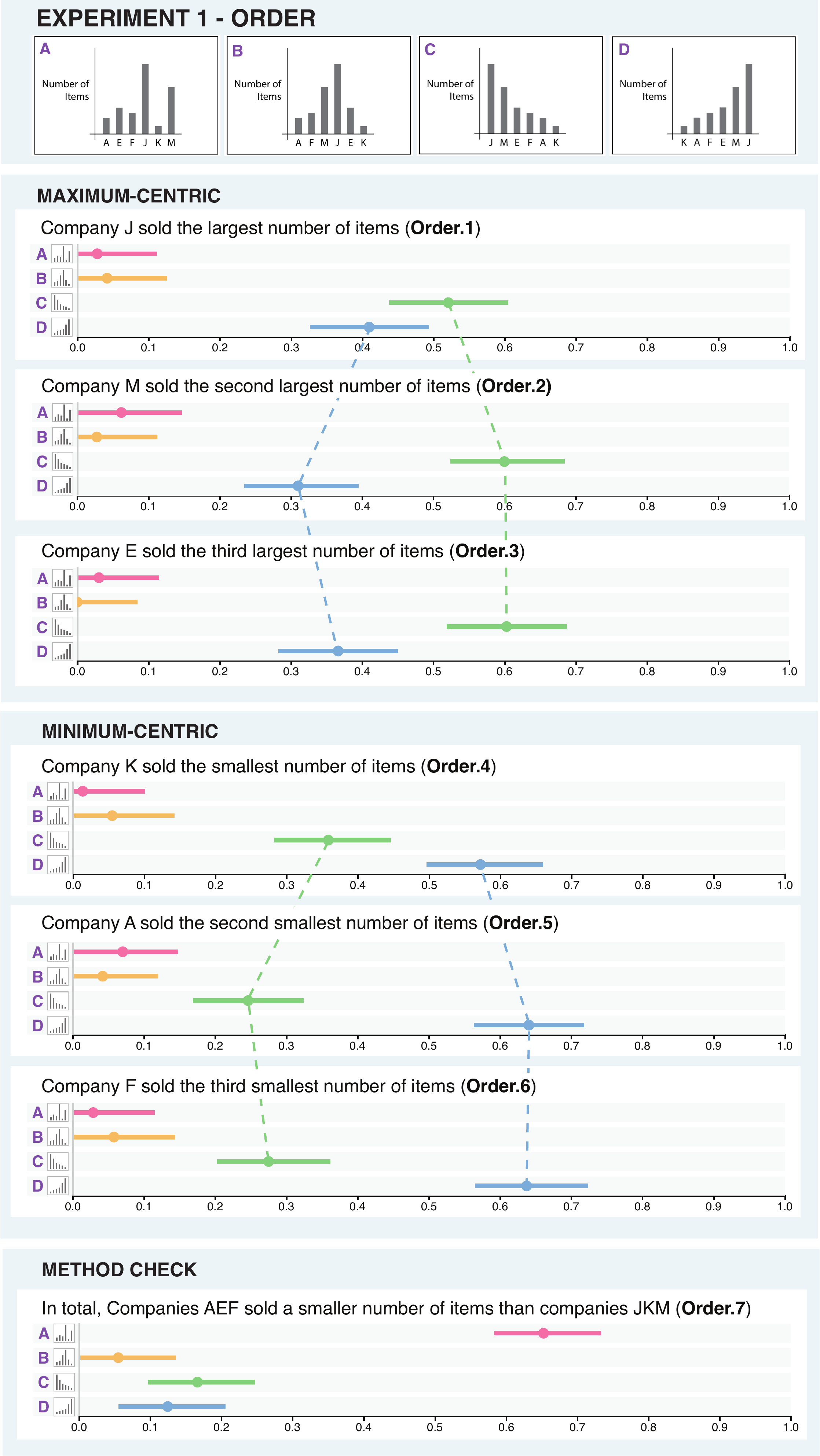}
\caption{Experiment 1 Results. Tested conditions are shown across the top of the figure. Below, lines encode 95\% CIs for the proportion of respondents that report each condition makes the given message most obvious. Circles encode the actual proportion observed from the experiment.}
\label{fig:exp1results}
\vspace*{-7mm}
\end{figure}

\begin{figure}[t!]
\includegraphics[width =0.49\textwidth]{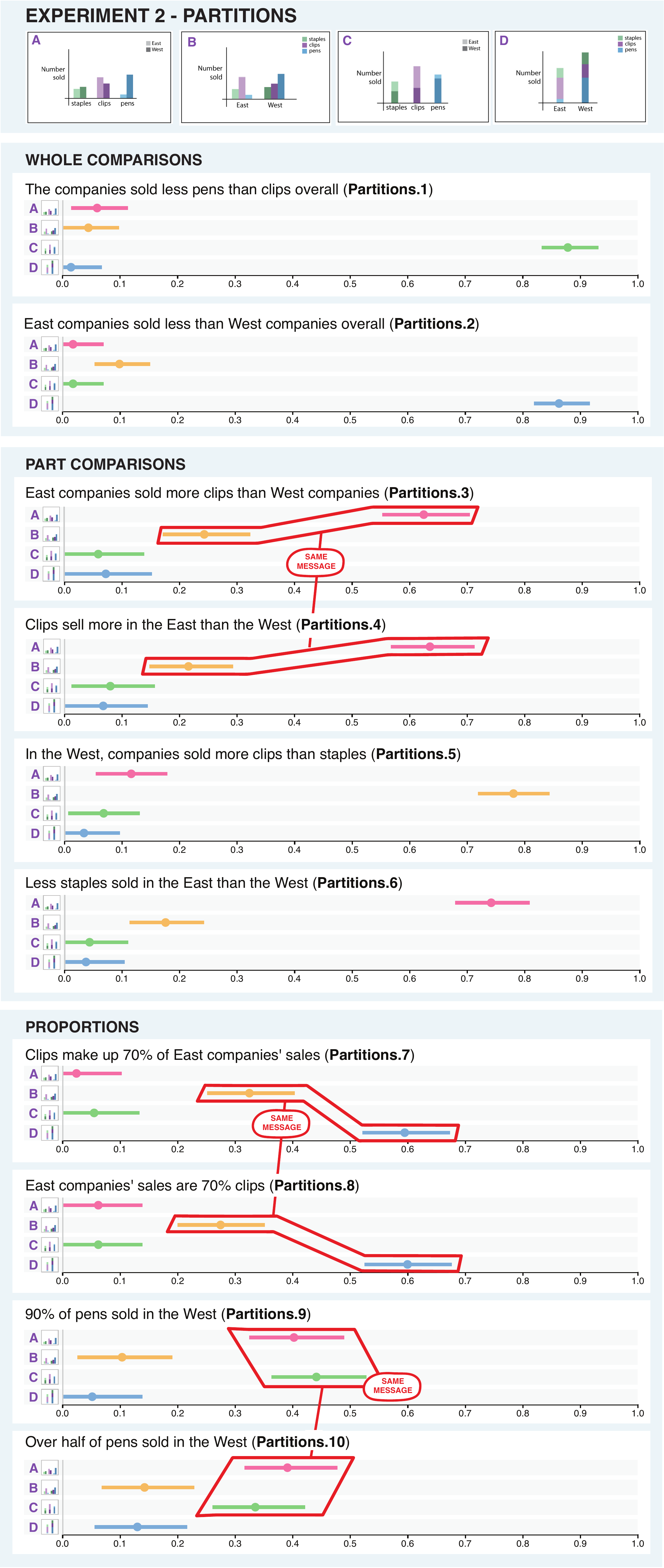}
\caption{Experiment 2 Results. Tested conditions are shown across the top of the figure. Below, lines encode 95\% CIs for the proportion of respondents that report each condition makes the given message most obvious. Circles encode the actual proportion observed from the experiment.}
\label{fig:exp2results}
\vspace*{-5mm}
\end{figure}

In \Cref{fig:exp1results,fig:exp2results,fig:exp3results,fig:exp4results}, we visualize the actual proportions and 95\% confidence interval for all four conditions given each message tested. We highly encourage all readers to view and determine strength of results for themselves, but will summarize visual findings using hedged language as advised by \cite{Dragicevic2016}.

\subsection{Experiment 1 Results}
Experiment 1 results are visualized in \Cref{fig:exp1results} and present strong support for hypotheses \textbf{H1A}, and \textbf{H1C} (H1B was rendered irrelevant in Exp. 1's re-run and thus dropped). For an overview of all hypotheses see \hyperref[sec:hyps]{Section 4.1}. 

The data in \Cref{fig:exp1results}, Maximum-Centric provide a consistent, visually distinct signal that bar charts formatted in descending order (condition C) make messages concerning the largest, second-, and third-largest bars more obvious than those formatted in ascending, alphabetical, or centrally-peaked order (conditions D, A, B). It is worth noting that this signal is much weaker for message Order.1, which concerns the largest value in the bar chart. Observed alone, there is not a sizeable difference in CIs to support condition C affording Order.1 more than condition D. Yet, when taken into context with messages Order.2 and Order.3, a more consistent signal of interest emerges.

The data in \Cref{fig:exp1results}, Minimum-Centric provide a similar series of visually distinct signals that bar charts formatted in ascending order (condition D) make messages concerning the smallest, second-, and third-smallest bars more obvious than those formatted in descending, alphabetical, or centrally-peaked order (C, A, B). This signal can be seen to grow stronger (i.e., the distance between CIs for ascending and descending conditions increases) as the messages concern more convoluted (e.g., second- and third- order) rankings. This increase in signal suggests that readers do not simply report increased obviousness due to marks of interest being immediately proximate to the left side of a chart, and that ascending and descending conditions still have an impact on obviousness of messages concerning bars that are more centrally located (e.g., third-largest and -smallest bars).

Finally, the data shown in \Cref{fig:exp1results}, Method Check support \textbf{H1C} with a strong signal that bar charts formatted in a particular order (condition A) make messages comparing companies grouped in that order (Order.7) more obvious than any other tested bar charts. This result is supported by cognitive psychology research that credits proximity with the ability to suggest grouping \cite{Wertheimer-1923-gestalt, century-gestalt-vis-perception, munzner2015visualization, ware-visual-thinking-for-design, brooks-chapter-perceptual-grouping, oyama-1961-proximity}. 

\subsection{Experiment 2 Results}
Experiment 2 results are visualized in \Cref{fig:exp2results}. Supporting \textbf{H2A}, the data in \Cref{fig:exp2results}, Whole Comparisons presents a consistent, visually distinct signal that, in part-to-whole charts, stacking bars (conditions C, D) make messages concerning comparison of the whole more obvious than arranging them side-by-side(conditions A, B).
Inversely, the data in \Cref{fig:exp2results}, Part Comparisons support \textbf{H2C} by presenting a consistent, visually distinct signal that bars arranged side-by-side (conditions A, B) make messages regarding the comparison of single parts more obvious than their stacked equivalents (conditions C, D).

 The data in \Cref{fig:exp2results}, Proportions provide fairly strong evidence to support \textbf{(H2B)} stacking bars (condition D) makes messages regarding a single part as a percentage of a three-part whole (message Partitions.7, .8) more obvious than a side-by-side arrangement (condition B). At the same time, the visualized CIs in  \Cref{fig:exp2results}, Proportions provide no evidence to support \textbf{(H2D)} the same difference in signal when messages regard a single part as a percentage of a two-part whole (messages Partitions.9, .10).
This difference could be explained by a visual processing capacity limit of two colors at once \cite{scimeca-tracking-muli-objects, xu-top}. For further discussion, see \Cref{sec:maintakeaways}.

Finally, Experiment 2 renders very similar CI results when testing re-wordings of the same messages (see red annotations in \cref{fig:exp2results}). This similarity supports \textbf{H2E} and the methodological validity of the survey by addressing concerns of potential confounding due to phrasing variations. 

\begin{figure}[t!]
\includegraphics[width=0.49\textwidth]{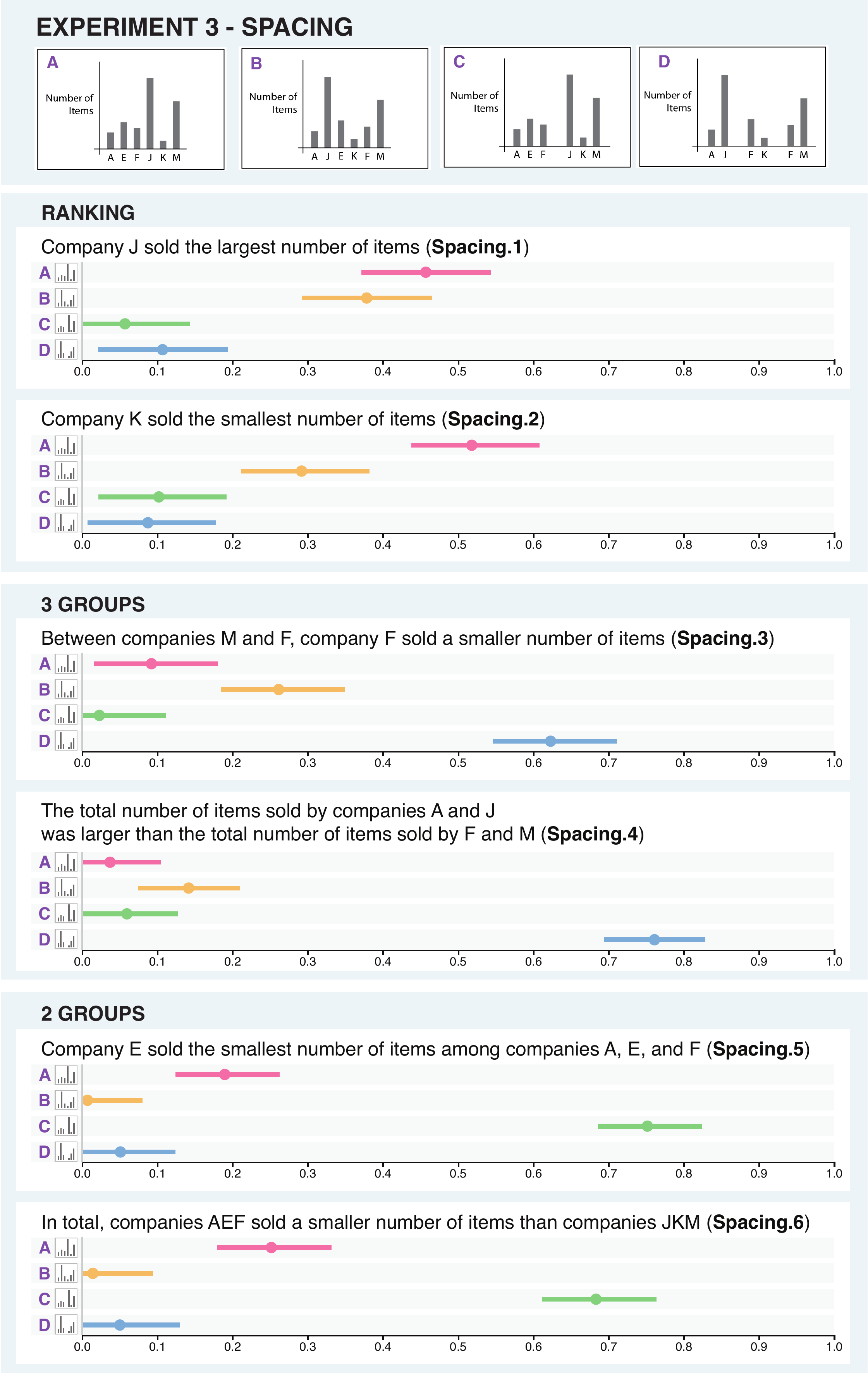}
\caption{Experiment 3 Results. Tested conditions are shown across the top of the figure. Below, lines encode 95\% CIs for the proportion of respondents that report each condition makes the given message most obvious. Circles encode the actual proportion observed from the experiment.}
\label{fig:exp3results}
\vspace*{-6mm}
\end{figure}

\subsection{Experiment 3 Results}
Experiment 3 results are visualized in \Cref{fig:exp3results}. The data in \Cref{fig:exp3results}, Ranking support \textbf{H3A} by depicting a consistent, visually distinct signal that bar charts without irregular spacing (conditions A, B) make messages concerning overall extrema more obvious than bar charts with irregular spacing (conditions C, D).
The data in \Cref{fig:exp3results}, 3 Groups and Figure 5, 2 Groups support \textbf{H3B} by displaying a consistent, visually distinct signal that bar charts grouped via irregular spacing (conditions C, D) make messages concerning those groups more obvious than bar charts with identical ordering but uniform spacing (conditions A, B).

\begin{figure}[t!]
\includegraphics[width=0.49\textwidth]{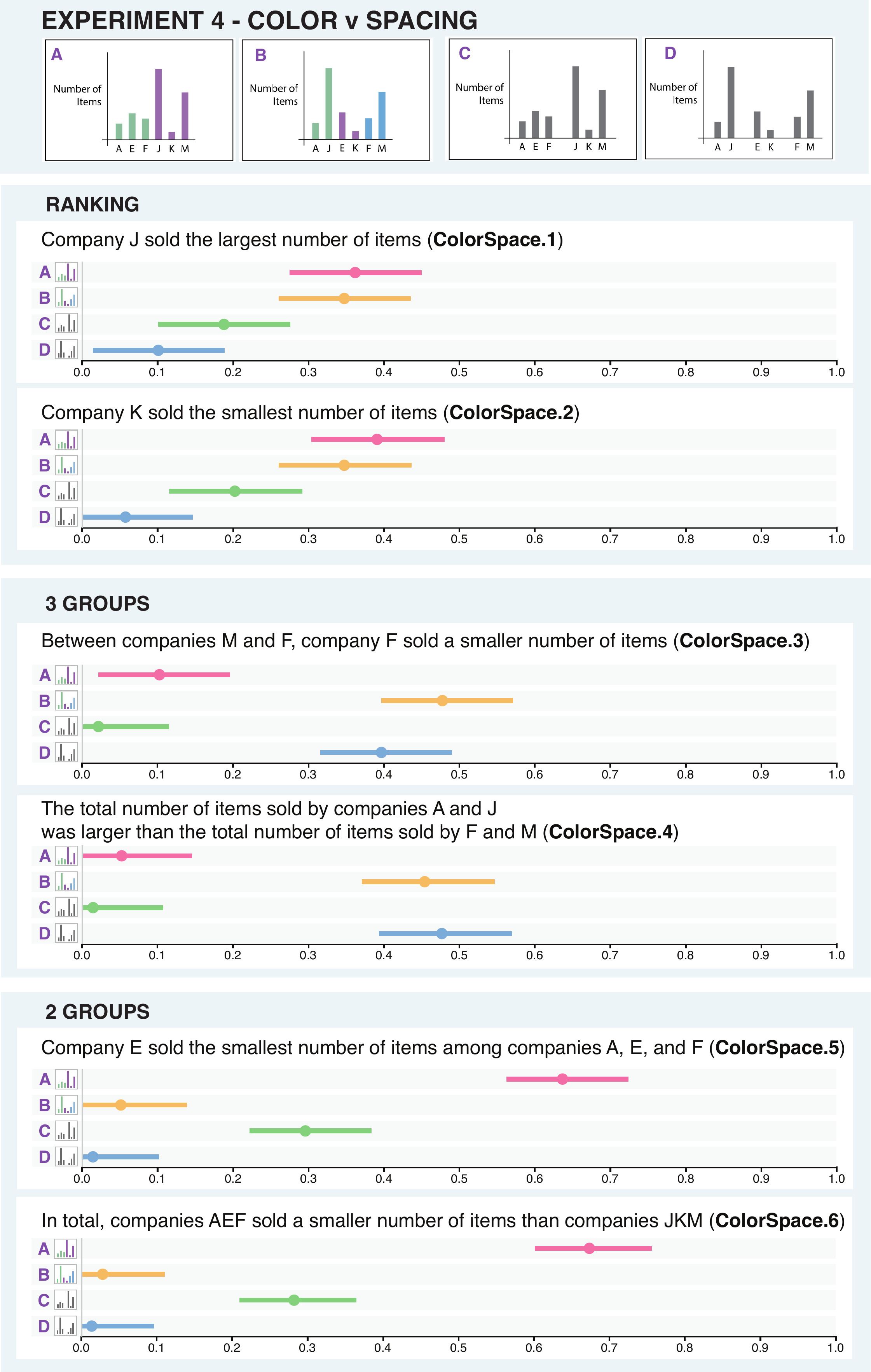}
\caption{Experiment 4 Results. Tested conditions are shown across the top of the figure. Below, lines encode 95\% CIs for the proportion of respondents that report each condition makes the given message most obvious. Circles encode the actual proportion observed from the experiment.}
\label{fig:exp4results}
\vspace*{-5mm}
\end{figure}

\subsection{Experiment 4 Results}
Experiment 4 results are visualized in \Cref{fig:exp4results}. The data in \Cref{fig:exp4results}, Ranking slightly support \textbf{H4A} with a consistent signal that bar charts with color grouping (conditions A, B) make messages concerning overall extrema more obvious than bar charts with proximity grouping (conditions C, D). The difference in signal between conditions A-B and C-D appear to be significant but are not as widely spread as hypothesis \textbf{H4A} postured.
The data in \Cref{fig:exp4results}, 2 Groups do not support \textbf{H4B}. Instead, they show a consistent, visually distinct signal that bar charts with color grouping (condition A) make messages concerning two groups of three bars more obvious than bar charts with spatial grouping (condition C), which is the inverse of our hypothesized hierarchy. The data in \Cref{fig:exp4results}, 3 Groups display visually approximate confidence intervals for conditions B and D, which--while different from results in \Cref{fig:exp4results}, 2 Group--still do not support \textbf{H4B}. We speculate the reason for this difference could be a visual processing capacity limit of two colors \cite{scimeca-tracking-muli-objects, xu-top}, as discussed in Experiment 2, as well. For further discussion, see \Cref{sec:maintakeaways}.

\section{Discussion}
\label{sec:discussion}
In this paper, we present four experiments investigating differences in visual arrangements' afforded messages. We do so through an empirical methodology that evaluates which arrangements of marks increase the obviousness of potential takeaways.

\subsection{Main Takeaways}
\label{sec:maintakeaways}
We summarize the following outcomes from the \hyperref[sec:results]{Results} section:

\begin{enumerate}
    \item Messages concerning largest, second-, and third- largest bars are made the most obvious by bars sorted in descending order from left to right (\cref{fig:exp1results}, Maximum-Centric).
    \item Messages concerning smallest, second-, and third- smallest bars are made the most obvious by bars sorted in ascending order from left to right (\cref{fig:exp1results}, Minimum-Centric).
\end{enumerate}
Takeaways 1 and 2 advise researchers and designers alike that the ordering of marks in a bar chart affects the affordance of messages about ranking. These takeaways exist within the context of the tested charts in Experiment 1 and the English-speaking nature of our participants. Still, these findings help bolster empirical evidence surrounding the impact of sorting bars, much of which has been conflicting. For example, Tversky et al. found similar evidence in their cognitive psychology study in 1991, discovering that children that speak directionally-ordered languages (e.g. left-to-right for English) associate ordering schema accordingly\cite{Tversky_Ordering}. %
At the same time, newer perceptual effort models, supported by eye-tracking experiments, suggest the opposite; ascending bars require more effort to extract the minimum value than descending bars\cite{schwartz-perceptual-effort-sorted-bars}. Regardless, this paper presents actionable recommendations for visualization designers who aim to draw attention to ranking-related messages.

\begin{enumerate}
  \setcounter{enumi}{2}
  \item In bar charts depicting part-to-whole data, messages concerning the whole(s) are made more obvious by stacking than by side-by-side arrangements (\cref{fig:exp2results}, Whole Comparisons).
   \item In bar charts depicting part-to-whole data, messages concerning the parts(s) are made more obvious by side-by-side arrangements than by stacking (\cref{fig:exp2results}, Part Comparisons).
   \item In bar charts depicting part-to-whole data, messages concerning parts as percentages of the whole are sometimes made more obvious by stacking (\cref{fig:exp2results}, Proportions).
\end{enumerate}
 
Takeaway 3 is hardly surprising since comparing visualized sums is easier than trying to mentally sum parts and then compare. The same, but inverse logic holds for Takeaway 4. More interestingly, Takeaway 5 finds messages about parts as a percentage of a whole are made equally, if not more, obvious by stacked bars over side-by-side bars. This holds true, even when an side-by-side arrangement would, from a precision standpoint, more effectively facilitate said comparison over its stacked counterpart \cite{heer-bostock-repr, talbot-bars}. Thus, we present initial evidence that precision and affordance (at least in the way it is operationalized as "obviousness" in our experiments) can diverge. In other words, a graph may lead to more precise comparisons and still be worse from an arrangement-message matching standpoint.

\begin{enumerate}
  \setcounter{enumi}{5}
  \item Spacing bars such that groups are formed by proximate bars makes messages concerning groupings of those bars more obvious than spacing bars uniformly. (\cref{fig:exp3results}, 3 Groups, 2 Groups)
   \item Uniformly spaced bar charts make messages concerning overall extrema more obvious than charts with irregular spacing. (\cref{fig:exp3results}, Ranking)
\end{enumerate}
Takeaway 6 is to be expected, as it is supported by decades of research on perceptual grouping\cite{Wertheimer-1923-gestalt, oyama-1961-proximity, oyama-1999-similarity-motion, brooks-chapter-perceptual-grouping, munzner2015visualization, ware-visual-thinking-for-design}. Thus, Takeaway 6 provides for a methodological sanity check. More interestingly, Takeaway 7 is  reiterative of the same perception research, but provides evidence that perceptual grouping can \textit{hinder} the affording of messages concerning groups. 
This result is also an interesting case in which precision-centric visualization guidelines do not align with affordances.  All arrangements in \hyperref[fig:exp3results]{Experiment 3} support the same level of precision when identifying a maximum.  Yet, our studies find bar charts with regular spacing make messages about maxima more obvious.

\begin{enumerate}
  \setcounter{enumi}{7}
   \item Uniformly spaced, color-grouped bar charts obstruct the obviousness of messages concerning extrema less than uncolored charts with spatial grouping (\cref{fig:exp4results}, Ranking).
   \item Uniformly spaced, color-grouped bar charts depicting two groups of three bars make messages concerning those groups more obvious than corresponding uncolored charts with spatial grouping \cref{fig:exp4results}, 2 Groups). Though this difference in obviousness is not apparent when the bar charts depict three groups of two (\cref{fig:exp4results}, 3 Groups). 
\end{enumerate}
Takeaway 8 aligns with Takeaway 7 and the current hierarchy of perceptual grouping techniques, which maintains that proximity more strongly indicates grouping than color\cite{yu-gestalt-not-parallel, yu-simi-grouping, benav-similarity-proximity, brooks-chapter-perceptual-grouping, munzner2015visualization}. As we see in Takeaway 7 from Experiment 3, grouping implied via proximity can hinder the obviousness of messages concerning extrema across groups. Thus, it is expected that color grouping would inhibit the obviousness of such messages slightly less than spatial grouping. 

Takeaways 5 and 9 were not predicted, and were surprising at first. However, after considering these takeaways in relation to recent accounts of the mechanisms underlying color grouping from perceptual psychology literature, we are excited that these findings might have a clear explanation, and with further testing, could result in clear and novel design guidelines. 

We first note that the human visual system can encounter a powerful capacity limitation when required to process multiple colors at the same time; many tasks, like grouping sets of objects by color, may even be a strictly serial process in which only one group is conceived in any one perceptual moment \cite{yu-gestalt-not-parallel, yu-simi-grouping}. In tasks that require people to temporarily associate a color with a label or other meaning (e.g., associating a color to a legend), capacity appears to be limited to two colors \cite{scimeca-tracking-muli-objects, xu-top}. 

The results associated with Takeaways 5 and 9 might be explained by this reluctance to process more than two colors at once. Recall that in the \hyperref[fig:exp2results]{Partitions experiment}, 'Proportions' comparisons involving pens in the West vs. the East (messages Partitions.9, .10) were rated as equally obvious for stacked and side-by-side bars. Note that this comparison should be between two bars, requiring inspection of two colors (light blue and dark blue). But comparisons of clips to two other products (messages Partitions.7, .8) require juggling three colors (that are also categorically different hues: purple, green, and blue), which may prove more aversive. In this three-color condition, participants suddenly strongly report that a stacked bar makes proportional messages the most obvious. In this case, the stacked bar might allow viewers to more easily select the relevant hue (purple) as a percentage of the whole bar, leading to a preference for arrangements that make information more clear when color capacity is reached.

Similarly, in \hyperref[fig:exp4results]{Experiment 4}, participants surprisingly preferred making comparisons between two groups of bars when those bars were defined by two colors instead of by two spatially separated regions. But when those groups were defined by three colors, participants were equivocal in their preference between color and spatial grouping. This finding could also be explained by an aversion to processing more than two colors at once. 

This explanation is speculative, and requires additional empirical support. Currently our studies confound number of colors, number of objects to be compared, and number of total groups. Additionally, some conditions use saturation differences (e.g., light purple and dark purple) while others use hue differences. While we doubt that these factors drive  asymmetries in our results, our understanding could be better supported by experiments that are specifically designed to isolate the effect of number of color hues.
Still, we remain excited about this speculative account because, while surprising, it is consistent with new models of color grouping and processing capacity\cite{scimeca-tracking-muli-objects, xu-top}. If this speculative reasoning holds, it would produce a clear design guideline: use color to distinguish among two groups, use either for three categories, and use space to distinguish among four or more.

\subsection{Limitations}
\label{sec:limitations}

While the method with which we study visualization affordances presents many positive features (see \Cref{sec:expapp}), its confirmatory nature also restricts the scope of possible findings. As noted in the \hyperref[sec:m&m]{Material \& Methods} section, our findings must be digested with their restricted scope in mind. For example, Takeaway 2 (\textit{Messages concerning smallest... bars are made the most obvious by bars sorted in ascending order from left to right.}) holds in comparison to the other three orders tested, but it may not do so when compared to other bar chart arrangements. Fortunately, this limitation can be mitigated in part by pairing our experimental design with an exploratory method, as detailed in the \hyperref[sec:relwork]{Related Work} section. %

 Similar contextual restrictions surround our study population. We recruit participants who are fluent in English, over the age of 18, and currently reside in the US. Ordered language conventions could very likely influence findings\cite{Tversky_Ordering}, and the replication of our work with other populations is prudent before generalizing results on a global scale. Fortunately, due to the easily replicable and modifiable nature of our method, such experiments could be run affordably.

Additionally, the results we present only consist of responses from participants who correctly filled in the drop-down of a given message. If a participant incorrectly filled in message A, their response to which chart made message A the most obvious was discarded, though all of their other responses were included. This exclusion has the potential to bias results towards an audience with a high graphic literacy. But to maintain a high quality of data, such removal is necessary to ensure that analyzed participants are actually answering survey questions with care. Due to both of these considerations, we provide a comparison of all results with and without this exclusion in SM4 in the Supplemental Materials -- no large differences are apparent between the two.

Lastly, while we posit that affordance is an important metric in evaluating visualizations, the line between affordance and effectiveness is blurry. Can a graph make a desired message obvious but be ineffective? Or can a graph be effective but not make a desired message obvious? These are questions that need to be clarified, but are difficult due to a lack of agreed upon definition for effectiveness in visualization (see \hyperref[sec:relwork]{Related Work} for a summary of metrics). Our current intuition, advised by our presented findings, is that affordance should correlate with increased graph comprehension, reduced reading effort, and general viewer preference. Future work is warranted to investigate if strong arrangement-message matches lead to increased efficacy of a graph, perhaps through the use of response time and precision as Vessey’s cognitive fit model suggests \cite{vessey-cog-fit-theory} or via other metrics like cognitive effort and memorability. 

\subsection{Implications \& Future Work}
\label{sec:impl}
The studies we present firmly suggest that visual arrangements can directly impact the messages people perceive from a graph. That is, the various arrangements of identical marks in a graph can alter the strength of perceived messages. While our experiments cover a limited set of visual arrangements and messages, they point to a number of implications, and compel the expansion of this work. 

To continue to build out academic and practical understanding of the effect of arrangements on afforded messages, our work can be extended as follows:
\begin{itemize}  
 \item \textit{Study different arrangement variations}. Future works may maximize their impact by investigating properties that are generic enough to apply to a wide variety of visual representations. Candidate arrangements include: orientation, rotation, styling of negative marks, and visual linking through outlines or edges.
 
 \item  \textit{Study different message types}. We cover a small subset of potential messages afforded by visualizations. Further exploration of other messages could drastically expand our understanding of visualizations and what they communicate.
 
 \item \textit{Study different visualization types}. Future work could also examine the arrangements studied in this paper (or an extension of them) with new types of graphs. Candidates include: spacing or ordering in pie charts or tree maps, color grouping in scatter plots or choropleths, and ordering in Sankey diagrams.

 \item \textit{Study the relative strength of arrangements’ affordance of messages}. Our methodology provides continuous, as opposed to binary, output allowing researchers to investigate both whether arrangements afford a message, and also possible hierarchies of arrangements’ affordance (e.g., spacing affords grouping $>$ color affords grouping $>$ shape affords grouping), as demonstrated in Experiment 4.

\end{itemize}

Lastly, the work presented in this paper has relevant implications for practitioners. This work provides infrastructure to build a “library” of visual arrangements and their afforded message,s which designers could use to inform and evaluate their visualizations. Practically, a designer could begin either with a desired message to communicate, or with a set of visualizations they want to narrow down, and use our framework, or a repository of results from our framework, to better understand the implications behind their designs.

The same affordance library could be used as an evaluation tool to review existing visualizations. Existing designs could be evaluated so as to confirm that intended messages are conveyed strongly and, equally paramount, that unintended messages are not strongly communicated.

\section{Conclusion}
\label{sec:conclusion}
In this paper, we investigate how four different arrangements of marks -- ordering, partitioning, spacing, and coloring -- in bar charts afford messages on ranking, part-to-whole relationships, and grouping. We present an replicable, scalable, modifiable, confirmatory methodology for investigating arrangements of marks within visualizations and their relative impact on afforded messages.

In our \hyperref[sec:relwork]{Related Work}, we establish current methods of investigating visualization affordances and current understanding of bar charts to provide context for our findings. 
In our \hyperref[sec:discussion]{Discussion}, we summarize our findings into nine key takeaways which provide insight for visualization designers, researchers, and educators on the affordance of messages when considering spatial and color arrangements of marks. We then contextualize said findings, comparing them to the closest existing research.

In summary, we provide two useful contributions: 1) four experiments resulting in nine takeaways on how bar chart arrangements afford various messages and
2) the tools to continue this work through an easily scalable and modifiable method for evaluating visualization arrangements' impact on their afforded messages.

\section*{Supplemental Materials}
\label{sec:supplemental_materials}
All supplemental materials are available on OSF at \url{https://osf.io/bvy95/files/osfstorage}.
In particular, they include (1-2) screenshots of the Qualtrics survey for posterity, (3) a table showing sample size for each tested message, (4) a side-by-side visual comparison of results excluding and including participants who answered a specific question incorrectly (does not apply to participants fully excluded from studies), (5) raw data files and a runnable jupyter notebook with all analysis, (6) .csv files used in the visualized CIs for \Cref{fig:exp1results,fig:exp2results,fig:exp3results,fig:exp4results}, and (7) a comparison of our work to \cite{xiong-visual-arrangement-of-bars}.

\section*{Figure Credits}
\Cref{fig:notallscatter} is a partial recreation of figures that appear in \cite{cairobars}.

\acknowledgments{%
The authors wish to thank Laura South, Myrl Marmarelis, and Sydney Purdue for their advice on statistical analysis. This work was supported in part by
a grant from the National Science Foundation (Award \#2236644)%
}

\bibliographystyle{abbrv-doi-hyperref}

\bibliography{template}

\appendix %

\end{document}